\documentclass[11pt,twoside]{article}

\usepackage{asp2004}
\usepackage{graphicx}

\begin{document}
\title{A Young Open Cluster Surrounding V838 Monocerotis }   
\author{Howard E. Bond and Melike Af\c{s}ar}   
\affil{Space Telescope Science Institute, 3700 San Martin Drive, Baltimore, MD
21218, USA; Department of Astronomy and Space Sciences, Ege University,
35100 Bornova, \.{I}zmir, Turkey}    

\begin{abstract} 
During a program of spectroscopic monitoring of V838~Mon, we serendipitously
discovered that a neighboring 16th-mag star is of type~B\null. We then carried
out a spectroscopic survey of other stars in the vicinity, revealing two more
B-type stars, all within $45''$ of V838~Mon. We have determined the distance to
this sparse, young cluster, based on spectral classification and photometric
main-sequence fitting of the three B~stars. The distance is found to be
$6.2\pm1.2$~kpc, in excellent agreement with the geometric distance to V838~Mon
of 5.9~kpc obtained from {\it Hubble Space Telescope\/} polarimetry of the light
echoes. The cluster's age is less than 25~Myr.

The absolute luminosity of V838~Mon during its outburst, based on our distance
measurement, was very similar to that of M31~RV, an object in the bulge of M31
that was also a cool supergiant throughout its eruption in 1988. However, there
is no young population at the site of M31~RV.

It does not appear possible to form a nova-like cataclysmic binary system within
the young age of the V838~Mon cluster, and the lack of a young population
surrounding M31~RV suggests that the outburst mechanism does not require a
massive progenitor. These considerations appear to leave stellar-collision or
\hbox{-merger} scenarios as one of the remaining viable explanations for the
outbursts of V838~Mon and M31~RV.

\end{abstract}

\section{Introduction}

A variety of explanations for the outburst of V838~Mon  have been proposed, many
of them mutually exclusive (see recent summary in Tylenda \&  Soker 2006, and
the  proceedings of this conference). These explanations involve either
thermonuclear processes (an unusual nova-like outburst on a white dwarf, a
thermonuclear event in a massive star, or a helium shell flash in a post-AGB
star), or the release of gravitational energy (through stellar or planetary
mergers or collisions).

A possible new constraint on the nature of V838~Mon came from the discovery of a
B3~V companion to the star (Munari \& Desidera 2002; Wagner \& Starrfield 2002).
The companion is unresolved even at {\it Hubble Space Telescope\/} ({\it HST\/})
resolution. 

Another eruptive object has attracted attention as a possible analog of
V838~Mon. This is the ``M31 red variable," or ``M31 RV,'' which underwent
an outburst in mid-1988 that was remarkably similar to that of V838~Mon (Bond \&
Siegel 2006 and references therein), although not as well observed.  M31~RV
occurred in the nuclear bulge of the Andromeda Galaxy, and is thus at a known
distance. 

In this paper, we present our serendipitous discovery that V838~Mon is a member
of a small, young open cluster. We will use spectral classification and
photometry of the cluster members to determine a distance. We will also derive a
limit to the cluster's age, and will compare the stellar populations surrounding
V838~Mon and M31~RV\null. We will close with a brief discussion of the V838~Mon
progenitor object and some new constraints on the outburst mechanism that result
from our observations.

\section{Observations and Data Reduction}

\subsection{Spectroscopy}

We have been monitoring the spectroscopic development of V838~Mon since early
2003, using the SMARTS 1.5-m telescope at Cerro Tololo Interamerican Observatory
(CTIO) and its CCD spectrograph. Most of our observations have been obtained
with a setup (grating designation ``26/I'') yielding a FWHM resolution of
4.3~\AA\ and a wavelength coverage of 3530--5300~\AA\null.  

Our data are long-slit spectra, in which the slit length projected onto the sky
is about $6'$. It is not unusual for neighboring field stars to fall onto the
slit, but we were surprised when a 16th-mag star lying on the slit almost
directly east of V838~Mon proved, entirely serendipitously, to have a B-type
spectrum.  Although V838~Mon lies at a low galactic latitude ($\ell=217\fdg8$,
$b=+1\fdg0$), a B-type star as faint as 16th~mag would lie at the outskirts of
the Galactic disk. This makes its existence very unusual, especially when lying
within a few arcseconds of a star that is itself extraordinary. Adding this to
the fact that V838~Mon itself has an unresolved B~companion made it appear very
likely that the serendipitous star is at the same distance. This in turn raised
the possibility that there could be further faint early-type objects in the
field surrounding V838~Mon.

To investigate this possibility, we used the SMARTS 1.5-m telescope to obtain
spectra of several more stars in the immediate vicinity of V838~Mon. For these
exploratory observations we used a setup (grating designation ``13/I''), giving
a resolution of 17.2~\AA\ and coverage 3150--9375~\AA\null. Most of the
neighboring stars have proven to be unrelated foreground stars, but our
observations to date have disclosed two further 14th-15th mag B-type stars near
V838~Mon. All three of these early-type stars lie within $45''$ of the variable,
or within a projected separation of only 1.3~pc if the distance is $\sim$6~kpc
(see below). Thus there is little doubt that V838~Mon is accompanied by a
previously unrecognized sparse, young cluster.

Figure~1 presents the Hubble Heritage image of V838~Mon and the light echo, with
the three new B~stars circled.  (In addition, V838~Mon itself is circled.) The
picture illustrates that the cluster is not obvious in a direct image, and as we
will see below the reddened B-type cluster members and foreground F-G stars have
similar colors; hence spectroscopic observations are the only practical means
for identifying cluster members in this crowded, low-latitude field.

\begin{figure}[ht]
\begin{center}
\includegraphics[width=2.5in]{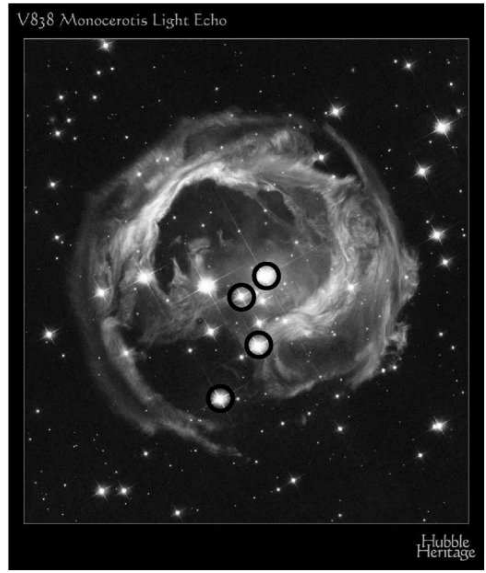}
\end{center}
\caption{Hubble Heritage image of V838~Mon and its light echo. The three
neighboring B-type stars discovered in our work are circled, along with V838~Mon
itself. From top to bottom, the circled stars are V838~Mon and stars 7, 9, and
8. Other bright stars in the image have proven to be foreground stars.}
\end{figure}

As this paper was being prepared, we became aware that Wisniewski, Bjorkman, \&
Magalh{\~a}es (2003) had already pointed out that our three
stars are likely to lie at a similar distance to V838~Mon itself, based on the
similar polarimetric properties of all four stars. 

\subsection{Spectral Classification}

Following the serendipitous discovery of the first field B-type star, and our
subsequent discoveries of two more B stars with the low-resolution 13/I grating
setup, we obtained moderate-resolution of the latter two with the SMARTS 1.5-m
spectrograph and the moderate-resolution 26/I grating. In Figure~2 we show these
spectra, along with spectra of several classification standards taken with the
same setup. Star~7 is the serendipitous star discovered first, and stars~8 and 9
are the other two B~stars. Also included in Figure~2 is a spectrum of V838~Mon
itself based on observations on five nights between 2003 February and May, a
time when the variable had declined considerably from its 2002 outburst, but was
still contributing some light even in this blue spectral range (note the TiO
bands longward of H$\beta$, for example).

\begin{figure}[t]
\begin{center}
\includegraphics[width=3.75in]{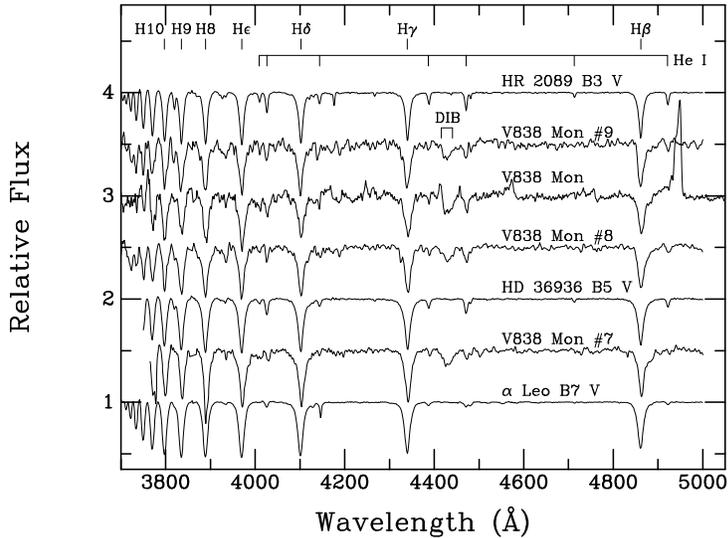}
\end{center}
\caption{SMARTS 1.5-m spectra of three B-type stars belonging to the young
cluster in the vicinity of V838~Mon, along with spectra of three classification
standards and V838~Mon itself (during early 2003). All spectra have been
normalized to a continuum level of 1.0, the tick marks on the $y$ axis are
separated by 0.5 continuum flux units, and the spectra have been offset by
constant successive amounts. Stars 9, 8, and 7 are classified B3~V, B4~V, and
B6~V, respectively, by direct comparison with the classification standards, as
described in the text. We also classify the unresolved companion of V838~Mon
itself as B3~V\null. Note the strong diffuse interstellar band (DIB) at
4428~\AA\ in the four reddened stars belonging to the V838~Mon cluster.}
\end{figure}

We classified the three B stars based both upon a visual comparison with the
standard stars shown in Figure~2, and upon equivalent-width measurements of the
He~I and Balmer lines. The resulting types are B3~V, B4~V, and B6~V,
respectively, for stars 9, 8, and 7. We were not able to  classify the
unresolved B-type companion of V838~Mon itself with these methods because of the
contamination from the cool component. However, from a comparison of the
equivalent widths of the bluest Balmer lines (H8, H9, and H10) in V838~Mon
(where the contamination is lowest) and in the standard stars, we find a type of
B3~V, which agrees well with the other authors cited in the introduction.

\subsection{Photometry}

We have also been monitoring the field of V838~Mon regularly since early 2003
with the SMARTS 1.3-m telescope and CCD direct camera. Calibrated photometry
of the stars surrounding V838~Mon on the Johnson $BV$ system was derived through
observations of a standard field from Landolt (1992).

\section{Distance and Other Properties of the Cluster}

\subsection{Spectroscopic Parallax}

We now calculate distances to each of the three B stars, and thus determine a
distance to the cluster, as shown in Table~1. The first four columns give the
star designations, spectral types, and photometry. The fifth column gives the
intrinsic color corresponding to each spectral type, $(B-V)_0$, taken from the
tabulation of Schmidt-Kaler (1982). The sixth column gives the estimated color
excess, $E(B-V) = (B-V)-(B-V)_0$. The color excesses agree well among the three
stars, with a mean of $E(B-V) = 0.84$ and a scatter of about $\pm$0.02~mag. This
determination agrees well with those of other authors. For example Munari et
al.\ (2005, hereafter M05) found $E(B-V) = 0.87\pm0.01$ using several different
methods. Tylenda (2005) discusses recent reddening determinations by several
authors, and adopts $E(B-V) \simeq 0.9$. 

\begin{table}[htb]
\caption{Spectroscopic Parallax Calculations}
\smallskip
\begin{center}
\small
\begin{tabular}{lcccccccc}
\tableline
\noalign{\smallskip}
Star &
Sp.\ &
$V$ &
$B-V$ &
$(B-V)_0$ &
$E(B-V)$ &
$V_0$ &
$M_V$ &
$(m-M)_0$ \\
\tableline
\noalign{\smallskip}
7 & B6 V & 16.02 & 0.71 & $-0.15$  &  0.86  &  13.42 &  $-0.9$ & 14.32 \\
8 & B4 V & 15.00 & 0.63 & $-0.19$  &  0.82  &  12.40 &  $-1.4$ & 13.80 \\
9 & B3 V & 14.79 & 0.62 & $-0.205$ &  \underline{0.825} &  12.19 &  $-1.6$ &
  \underline{13.79} \\         
  &   &  &	&	  & \llap{Mean: } 0.84 & & & 14.0 \\
\noalign{\smallskip}
\tableline
\end{tabular}
\end{center}
\end{table}

The next column in Table~2 gives the magnitude of each star corrected for
extinction, $V_0$, calculated assuming $A_V=3.1E(B-V)$\null. Column~8 gives the
absolute magnitude corresponding to each spectral type, $M_V$, again taken from
Schmidt-Kaler (1982). The final column gives the distance moduli, $(m-M)_0$,
whose mean is 14.0, or a distance of 6.3~kpc. 

The cluster distance modulus has an internal error of about $\pm$0.2~mag, based
on the scatter among the three stars. However, systematic errors are undoubtedly
larger. This is indicated by the scatter among different calibrations of the
relation between spectral types and absolute magnitudes (e.g., Lesh 1968;
Schmidt-Kaler 1982; Cramer 1997), which amounts typically to about $\pm$0.4~mag.
At a distance of 6.3~kpc, this corresponds to an error of $\pm$1.2~kpc.

\subsection{Main-sequence Fitting}

We can also estimate the distance to our cluster through photometric
main-sequence fitting. In the absence of any spectroscopic information, this
method would suffer from the well-known near-degeneracy between extinction and
distance for early-type stars; this is due to fact that the main sequence lies
along a steep, nearly straight line in the $V,(B-V)$ diagram. However, with the
additional constraints from the spectral types, main-sequence fitting becomes
possible.

To define the main sequence, we chose the lightly reddened open cluster
NGC~2362, which, at an age of $\sim$5~Myr, has been described as a template for
early stellar evolution (Moitinho et~al.\ 2001; hereafter MAHL01). This
cluster's unevolved main sequence extends to type B1~V (Johnson \& Morgan 1953).
We took $B,V$ photometry for NGC~2362 from Johnson \& Morgan and from Perry
(1973), and adopted $E(B-V)=0.10$ and $d=1.48$~kpc from MAHL01. We then
corrected the photometry to the spectroscopic reddening and distance for the
V838~Mon cluster found above.  The match with the three V838~Mon B stars was
very good, so we applied just one iteration of adjusting first the reddening of
the B stars and then the distance, so as to improve the fit. This resulted in a
V838~Mon cluster reddening of $E(B-V)=0.85$ and distance modulus $(m-M)_0=13.97$
($d=6.2$~kpc).  The external errors here are similar to those for the
spectroscopic parallax, since MAHL01 based the distance to NGC~2362 on the
Schmidt-Kaler (1982) zero-age main sequence.

In Figure~3 we plot the $V,(B-V)$ values for the three B stars as large filled
circles.  The small open circles are the Johnson-Morgan and Perry
photometry of NGC~2362, adjusted to the V838~Mon reddening and distance found in
the previous paragraph. The fit is excellent, and strongly supports the
conclusion that our three B stars do form a physical cluster.

\begin{figure}[ht]
\begin{center}
\includegraphics[width=3.15in]{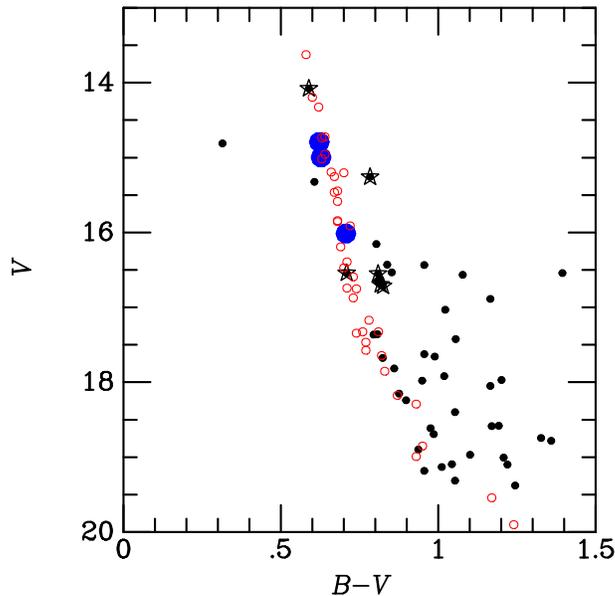}
\end{center}
\caption{Color-magnitude diagram for the three B stars belonging to the
V838~Mon cluster ({\it large filled circles}). Also plotted are all stars
within a $90''$ radius of V838~Mon ({\it small black filled circles}) for which
we do not have spectra, except for six spectroscopically confirmed foreground
field stars ({\it black star symbols}). The {\it open circles\/} plot
photometry for the template zero-age main sequence of the open cluster NGC~2362
(Johnson \& Morgan 1953; Perry 1973), adjusted to a reddening of $E(B-V)=0.85$
and a distance of 6.2~kpc; field stars and binaries have been omitted.}
\end{figure}

A direct geometrical distance determination for V838~Mon has been carried out by
Sparks et al.\ (2006), based on polarimetric imaging of the light echo obtained
with {\it HST\/}\null. Their result, 5.9~kpc, is in excellent agreement with our
determination of 6.2~kpc based on the associated B-type stars.

Also plotted in Figure~3, as small filled black circles, is our photometry for
all stars within a radius of $90''$ of V838~Mon. For most of these stars we do
not know whether they are cluster members or not, but we do have SMARTS 1.5-m
spectra for six of them that establish them as belonging to the foreground;
these non-members are marked with black stars in Figure~3.

As Figure~3 shows, the foreground F- and G-type stars have similar colors to the
reddened B-type cluster members. This demonstrates that spectroscopic
observations are essential to the identification of cluster members. It would be
very interesting to have spectra of the other stars in the vicinity of V838~Mon,
especially below $V\simeq17.5$, where the slope of the main sequence changes;
any of these candidates that proved to be cluster members would provide tighter
constraints on the cluster reddening and distance.

We can compare the luminosity of V838~Mon with that of the apparently similar
object M31~RV (see \S1). At maximum light (2002 February 6) V838~Mon had $B=7.9$
(M05, their Figure~1). For the reddening and distance derived here, this
corresponds to an absolute blue magnitude at maximum of $M_B=-9.6$. The
brightest $B$ magnitude measured for M31~RV during its 1988 outburst was 17.3
(Bryan \& Royer 1992; Boschi \& Munari 2004). The latter authors, adopting
$E(B-V)=0.12$ and $(m-M)_0=24.48$ for M31, found an absolute magnitude of
$M_B=-7.7$.  However, the light curve of M31~RV was very poorly sampled around
its maximum. It is more meaningful to compare the luminosities of the two
objects at the same well-covered portions of both light curves. Referring to
Figure~2 of Boschi \& Munari (2004), and using their light curves in
Kron-Cousins $R$, we can compare the luminosities at a point just before the
rapid fading in the $R$ band.  For M31~RV, this brightness is $R\simeq15$, and
for V838~Mon it is $R\simeq6.2$.  Correcting for reddening and distance, the
corresponding absolute magnitudes are $M_R\simeq-9.8$ and $-9.7$, respectively.
Thus, at least at this stage in their outbursts, the absolute luminosities were
nearly identical.

\subsection{Cluster Age and Stellar Masses}

Since the three B stars all appear to lie on the zero-age main sequence, we can
only set an upper limit to the age of the V838~Mon cluster. By reference to the
isochrones and evolutionary tracks of Pietrinferni et al.\ (2006), we find
that this upper limit is about 25~Myr. 


\section{The Stellar Populations of V838 Mon and M31 RV}

V838~Mon is accompanied by a previously reported, unresolved B3~V companion. In
this paper we have shown that V838~Mon also belongs to a sparse cluster,
containing at least three other B-type members.

We could thus be tempted to speculate that the outburst of V838~Mon in 2002
represents an explosive event that occurs in stars of masses
$\ga$7--$8\,M_\odot$\null.  Unfortunately, such a speculation appears to be
dashed by the recent study of M31~RV by Bond \& Siegel (2006). They used
archival {\it HST\/} images that serendipitously included the outburst site of
M31~RV, and showed that the population surrounding the object contains
exclusively old, low-mass stars belonging to the nuclear bulge of M31. There is
no young population at all at this site, let alone bright B stars younger than
25~Myr.

Thus, if the outbursts of both V838~Mon and M31~RV arose from a common
mechanism, the mechanism occurs among both very young and very old
stars.   

\section{The Outburst Mechanism}

Our observations have revealed that V838~Mon belongs to a young cluster. This
discovery has, however, only deepened the enigma of V838~Mon and the similar
object M31~RV, because we now know that the former arose from a very young
population, whereas the latter belongs to a very old population.

What, then, was the nature of the progenitor objects that produced the outbursts
of V838~Mon and M31~RV\null? They apparently can exist at both ends of the range
of stellar ages. The luminosity of the progenitor before outburst, at least in
the case of V838~Mon, was small compared to that of a B3 dwarf (see Af\c{s}ar \&
Bond 2006). That would be consistent with the low luminosity of a typical
nova-like cataclysmic variable. However, the young age of the V838~Mon cluster
would appear to rule out the evolutionary timescale required to produce an
accreting white dwarf in a compact binary, a requirement of a nova-like outburst
mechanism. 

It may be that the stellar-merger scenario advocated by several participants at
this conference, as well as in a recent series of
papers (e.g., Tylenda \& Soker 2006 and references therein; see also Retter et
al.\ 2006, who advocate a planet-star merger) can satisfy these new constraints,
provided that collisions between low-mass stars (which exist in all populations)
could produce the required high luminosities, and can be shown to occur often
enough. At the moment, however, the nature of these extraordinary outbursts
remains one of the leading unsolved problems in stellar astrophysics.

\acknowledgements 

M.~A. would like to thank The Scientific and Technological Research Council of
Turkey (TUBITAK) and her advisor Prof.\ Cafer \.{I}bano\v{g}lu, as well as
H.E.B. and STScI for their support during her doctoral studies. STScI's
participation in the SMARTS consortium is supported by the STScI Director's
Discretionary Research Fund. We thank numerous colleagues for discussions of
V838~Mon, including especially the members of the {\it HST\/} V838~Mon observing
team, and the participants in this conference.

\end{document}